\documentclass[twocolumn,groupeaddress,showpacs,aps,pra]{revtex4}

\usepackage{hyperref}
\usepackage{xcolor}
\usepackage{url}

\usepackage{amsmath}
\usepackage{amsfonts}
\usepackage{amssymb}
\usepackage{bm}
\usepackage{bbm}
\usepackage{nicefrac}
\usepackage{xcolor}
\usepackage{pifont}
\usepackage{graphicx}
\usepackage{enumerate}
\usepackage{dcolumn}
\usepackage{maybemath}

\newcommand{\ee}{\mathrm{e}}

\newcommand{\calF}{{\mathcal F}}
\newcommand{\calL}{{\mathcal L}}

\definecolor{oxfordblue}{rgb}{0.05, 0.2, 0.7}
\definecolor{cambridgeblue}{rgb}{0.1, 0.3, 1.0}

\def\lambdabar{{\mskip2mu\mathchar'26\mkern-9.8mu\lambda}}

\begin{document}

\title{Muonic bound systems, virtual particles and proton radius}

\author{U. D. Jentschura}

\affiliation{Department of Physics,
Missouri University of Science and Technology,
Rolla, Missouri 65409, USA}

\begin{abstract}
The proton radius puzzle questions the self-consistency of theory and
experiment in light muonic and electronic bound systems. Here, we summarize the
current status of virtual particle models as well as Lorentz-violating
models that have been proposed in order to explain the discrepancy.  Highly
charged one-electron ions and muonic bound systems have been used as probes of
the strongest electromagnetic fields achievable in the laboratory.  
The average electric field seen by a muon orbiting a proton is comparable
to hydrogenlike Uranium and, notably, larger than the electric field in the
most advanced strong-laser facilities. Effective interactions 
due to virtual annihilation inside the proton (lepton pairs)
and process-dependent corrections (nonresonant effects) are discussed as
possible explanations of the proton size puzzle.  The need for more
experimental data on related transitions is emphasized.
\end{abstract}

\pacs{12.20.Ds, 11.25.Tq, 11.15.Bt}

\maketitle

%
%
\section{Introduction}

Recent muonic hydrogen experiments~\cite{PoEtAl2010,AnEtAl2013} have resulted
in the most severe discrepancy of the predictions of quantum electrodynamics
with experiment recorded over the last few decades.  In short, both
(electronic, i.e., atomic) hydrogen experiments (for an overview see
Ref.~\cite{JeKoLBMoTa2005}) as well as recent scattering experiments lead to
a proton charge radius of about $\left< r_p \right> \approx 0.88 \, {\rm
fm}$, while the muonic hydrogen experiments~\cite{PoEtAl2010,AnEtAl2013}
favor a proton charge radius of about $\left< r_p \right> \approx 0.84 \,
{\rm fm}$.

Few-body bound electronic and muonic systems belong to the 
most intensely studied fundamental physical entities;
a combination of atomic physics and field-theoretic
techniques is canonically 
employed~\cite{BeSa1957,BrMo1978,BeKlSh1997,SoEtAl1998,MoPlSo1998,%
SoEtAl2001,EiGrSh2001,EiGrSh2007,Je2011aop1,Je2011aop2,AnEtAl2013aop}.
Here, we aim to discuss conceivable explanations for the discrepancy and
highlight a few aspects that set the muonic systems apart from any other
bound states which have been studied spectroscopically so 
far.  To this end,
in Sec.~\ref{sec2}, we briefly summarize the status of virtual particle
models discussed in the literature and supplement previous approaches
with a discussion of the role of axion terms that might be significant in
the strong magnetic fields used in the muonic hydrogen experiments.  In
Sec.~\ref{sec3}, we show that muonic hydrogen (as well as muonic hydrogenlike
ions with low nuclear charge number $Z$) constitute some of the most
sensitive probes of high-field physics to date; concomitant speculations
about novel phenomena in the strong fields inside the proton are discussed.
Finally, a possible role of process-dependent corrections in experiments is
mentioned in Sec.~\ref{sec4}.  Conclusions are reserved for Sec.~\ref{sec5}.
We use SI mksA units unless indicated otherwise.

%
%
\section{Virtual Particles and Muonic Hydrogen}
\label{sec2}

From the point of view of quantum field theory,
the most straightforward explanation for the proton radius 
puzzle in muonic hydrogen would involve a 
``subversive'' virtual particle that modifies the 
muon-proton interaction at distances commensurate with the
Bohr radius of muonic hydrogen,
\begin{equation}
\label{amu}
a_\mu = \frac{\hbar}{\alpha_{\rm QED} \, m_r \, c} = 
2.84708 \times 10^{-13} \, {\rm m} \,,
\end{equation}
where $\alpha_{\rm QED}$ is the fine-structure constant
and $m_r = m_\mu \, m_p/(m_\mu + m_p)$ is the reduced mass.
The distance regime of $a_\mu \approx 300 \, {\rm fm}$ is 
intermediate between the Bohr radius of (ordinary) hydrogen 
and the proton radius.

In consequence, the possible role of millicharged 
particles, which modify the Coulomb force law in this 
distance regime, has been analyzed in Ref.~\cite{Je2011aop2}.
These particles could conceivably modify the photon
propagator at energy scales $\hbar c/a_\mu$ 
via vacuum-polarization insertions into the photon line.
Supplementing this analysis, in Ref.~\cite{Je2011aop1},
conceivable hidden (massive) photons are analyzed.
Particles with scalar and pseudo-scalar couplings 
are the subject of Ref.~\cite{CaRi2012}.
A model that explicitly  breaks electron-muon 
universality, introducing a coupling of the right-handed 
muonic fermion sector to a $U(1)$ gauge boson,
is investigated in Ref.~\cite{BaMKPo2011}.
One should notice, though, that the explicit breaking 
of the universality according to Eq.~(7) of Ref.~\cite{BaMKPo2011}
appears as somewhat artificial. The 
reduction in the muonic helium nuclear radius 
by $\Delta r^2_{\rm He} = -0.06 \, {\rm fm}^2$
as predicted by the model proposed in Ref.~\cite{BaMKPo2011}
has the opposite sign as compared to the results of the 
experiments~\cite{CaEtAl1969,BrEtAl1972}, that were carried out about four
decades ago and observe a roughly 4\,\% {\em lower} cross section for muons
scattering off of protons as opposed to electrons being scattered off the same
target. 

Likewise, in a recent paper on Lorentz-violating terms in effective Dirac
equations~\cite{GoKoVa2014}, the authors assume an explicit breaking of
electron-muon universality (see Sec.~IIC3 of Ref.~\cite{GoKoVa2014}
where the authors explicitly state that they
assume only muon-sector Lorentz violation, so that effects arise in H$_\mu$
spectroscopy but are absent in H spectroscopy and electron elastic scattering).
Viewed with skepticism, this assumption appears to be a little artificial
because it would modify the effective Dirac equation for muons as compared to
that of electrons.  In general, Lorentz-violating parameters may break
rotational invariance, and thus have an effect on the $S$--$P$ transitions
measured in~\cite{PoEtAl2010,AnEtAl2013} [see the derivation in Eqs.~(21)--(25)
of Ref.~\cite{GoKoVa2014}].

In the virtual particle models from
Refs.~\cite{Je2011aop2,CaRi2012,BaMKPo2011}, it has been found necessary to
fine-tune the coupling constants in order to avoid conflicts with muon and
electron $g-2$ measurements, which otherwise provide constraints on the size of
the new physics terms due to their relatively good agreement with
experiment (for a discussion, see Ref.~\cite{Je2011aop2}).  Furthermore,
attempts to reconcile the difference based on higher moments of the proton
charge distribution (its higher-order shape, see Ref.~\cite{dR2010}) face
difficulty when confronted with scattering experiments that set relatively
tight constraints on the higher-order corrections to the proton's shape.

One class of models that has not been explored hitherto concerns
electrodynamics with axion-like particles (ALPs, 
see Refs.~\cite{PeQu1977,GiJaRi2006prl,%
AhGiJaRi2007,AhGiJaReRi2008}).
In the experiments~\cite{PoEtAl2010,AnEtAl2013},
strong magnetic fields on the order of about $5 \, {\rm T}$
are used to collimate the muon beam.
Axion terms could potentially influence the results of the 
spectroscopic measurements.
We start from the Lagrangian~\cite{Si1983,Gi2008,EhEtAl2009,EhEtAl2010,PDG2012}
for a pseudoscalar ($0^-$) axion-like particle
(temporarily setting $\hbar = c = \epsilon_0 = 1$)
\begin{align}
\label{calL}
{\mathcal L} =& \; -\frac14 \, F^{\mu\nu} \, F_{\mu\nu}
- \frac{g}{4} \, \phi \, \widetilde F^{\mu\nu} \, F_{\mu\nu} 
\nonumber\\[0.133ex]
& \;
+ \frac{1}{2} \partial_\mu \phi \, \partial^\mu \phi - \frac{1}{2} m_\phi^2 \, \phi
\nonumber\\[0.133ex]
=& \; \frac12 \, (\vec E^2 - \vec B^2)
+ g \, \phi \, \vec E \cdot \vec B
\nonumber\\[0.133ex]
& \; + \frac{1}{2} 
\partial_\mu \phi \, \partial^\mu \phi - \frac{1}{2} m_\phi^2 \, \phi \,.
\end{align}
Here, according to Ref.~\cite{PDG2012},
the axion's two-photon coupling constant reads as
\begin{equation}
g \equiv G_{A\gamma\gamma} =
\frac{\alpha_{\rm QED}}{2 \pi f_A} \, 
\left( \frac{E}{N} - \frac23 \, \frac{4 + z}{1+z} \right)
\end{equation}
($\phi$ is the axion field,
$m_\phi$ is the axion mass, $m_\phi \, f_A \approx m_\pi f_\pi$ 
where $f_\pi$ is the pion decay constant and $m_\pi$ 
the pion mass, while $z = m_u/m_d$ is the quark mass ratio).
Grand unified models~\cite{DiFiSr1981,Zh1980,Ki1979,ShVaZa1980,ChGeNi1995}
assign rational fractions to the ratio $E/N$ of the 
electromagnetic to the color anomaly of the axial current 
associated with the axion.
Possible values are $E/N = 8/3$ (see Refs.~\cite{DiFiSr1981,Zh1980})
or zero~\cite{Ki1979,ShVaZa1980}.
In Eq.~\eqref{calL}, the electromagnetic field strength tensor 
$F_{\mu\mu}$ and its dual $\widetilde F_{\mu\nu}$ 
have their usual meaning. 

\begin{figure}[t!]
\begin{center}
\begin{minipage}{0.91\linewidth}
\includegraphics[width=0.5\linewidth]{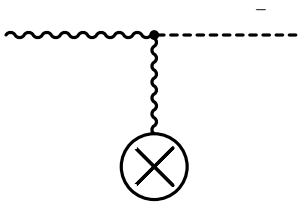}
\caption{\label{fig1} ALP-photon conversion in a 
strong magnetic field according to the interaction 
term in the Lagrangian given in Eq.~\eqref{calL}.
The large encircled cross denotes the interaction 
with an external magnetic field.}
\end{minipage}
\end{center}
\end{figure}

\begin{figure}[t!]
\begin{center}
\begin{minipage}{0.91\linewidth}
\includegraphics[width=0.6\linewidth]{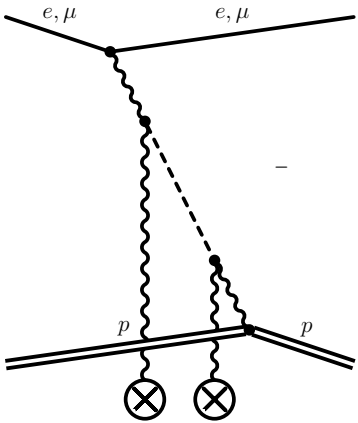}
\caption{\label{fig2} The leading (tree-level) correction to 
the Coulomb potential due to the ALP-photon interaction 
is given by the tree-level diagram shown.
The upper fermion line corresponds to an
electron $e$ (ordinary hydrogen) or a 
muon $\mu$ (muonic hydrogen).}
\end{minipage}
\end{center}
\end{figure}

It is interesting to consider the leading 
correction to the Coulomb potential in strong magnetic 
fields, on the order of $5 \, {\rm T}$,
due to the axion-photon conversion amplitude inherent to the 
Lagrangian~\eqref{calL} (see Figs.~\ref{fig1} and~\ref{fig2}).
We shall first assume that the vacuum expectation value of 
the axion field vanishes~\cite{VCdP2013,VC2014} and consider the tree-level correction
to the Coulomb potential given in Fig.~\ref{fig2}.

\begin{figure}[t!]
\begin{center}
\begin{minipage}{1.0\linewidth}
\includegraphics[width=0.91\linewidth]{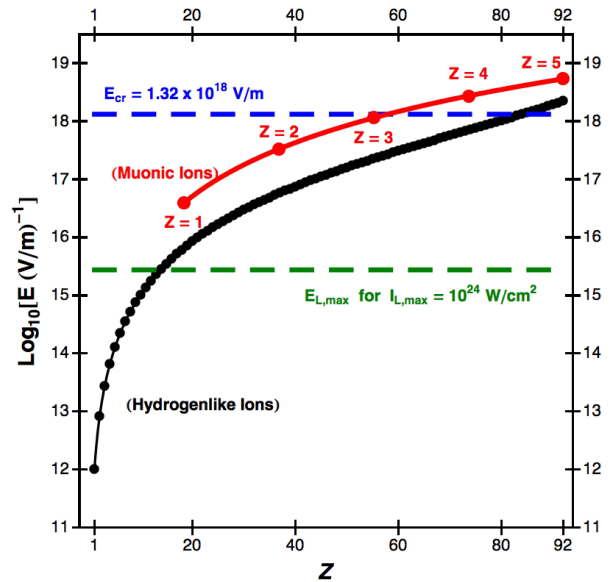}
\caption{\label{fig3} (Color.) Plot of the average field 
strength $E \equiv \langle E \rangle$ 
[see Eq.~\eqref{stda}] experienced by a bound electron or muon
in a one-muon ion (red line, $1 \leq Z \leq 5$),
and for hydrogenlike (electronic) ions 
in the range $1 \leq Z \leq 92$.
For comparison, the average field strength in a laser field 
of intensity $10^{24} \, {\rm W} \, {\rm cm}^{-2}$ 
is given~\cite{YaEtAl2008}.
The Schwinger critical field strength is
denoted by $E_{\rm cr}$.}
\end{minipage}
\end{center}
\end{figure}

We match the scattering amplitude 
according to Chap.~83 of Ref.~\cite{BeLiPi1982vol4}
(see also~\cite{CaLe1986}) and calculate the 
potential, generated by the axion-like particle,
due to the diagram in Fig.~\ref{fig2}.
The pseudoscalar ALP potential is given as
\begin{align}
V_{{\rm ALP} \, 0^-}(\vec k) =& \; (\vec k \cdot \vec B)^2 \;
\frac{4 \pi Z \alpha \, g^2}{\vec k^{\,4} \, (\vec k^2 + m_\phi^2)}
= (\vec k \cdot \vec B)^2 \; f(\vec k) \,,
\nonumber\\[0.133ex]
f(\vec k) =& \; \frac{4 \pi Z \alpha \, g^2}{\vec k^{\,4} \, (\vec k^2 + m_\phi^2)} \,.
\end{align}
In coordinate space, we therefore have
\begin{align}
V_{{\rm ALP} \, 0^-}(\vec r) =& \; 
- \left( \vec B \cdot \vec \nabla \right)^2  f(\vec r) \,,
\nonumber\\[0.133ex]
f(\vec r) =& \; 4 \pi Z\alpha g^2 \, \left(
\frac{\ee^{-m_\phi \, r} - 1}{4 \pi \, m_\phi^4 \, r}
- \frac{r}{8 \, \pi \, m_\phi^2} \right) \,.
\end{align}
With $f(\vec r) = f(r)$, we have
the second derivative as
\begin{equation}
\left( \vec B \cdot \vec \nabla \right)^2 f(r) =
\left( \frac{\vec B^{\,2}}{r} -
\frac{(\vec B \cdot \vec r)^2}{r^3} \right) f'(r) +
\frac{(\vec B \cdot \vec r)^2}{r^2} f''(r) .
\end{equation}
Differentiating and expanding for small $m_\phi$, one obtains
\begin{align}
V_{{\rm ALP} \, 0^-}(\vec r) =& \;
Z\alpha g^2 \, \left( \frac{\vec B^{\,2}}{3 m_\phi} -
\frac{ \vec B^2 \, \vec r^{\,2} + (\vec B \cdot \vec r)^2}{8 \, r} \right)
\nonumber\\[0.133ex]
\sim & \;  -\frac{Z\alpha g^2}{8 \, r} \, 
\left( \vec B^{\,2} \, \vec r^{\,2} + (\vec B \cdot \vec r)^2 \right)
\end{align}
where we subtract the constant shift.
This effective potential is independent of the ALP mass
$m_\phi$ provided $m_\phi$ is much smaller than other mass
scales in the problems, such as $m_e$ and $m_\mu$
(see also Fig.~\ref{fig3}).
The $1S$ expectation value is 
\begin{align}
\delta E = & \; \left< 1S \left| -\frac{Z\alpha g^2}{8 \, r} \,
\left( \vec B^{\,2} \, \vec r^{\,2} + (\vec B \cdot \vec r)^2 \right) 
\right| 1S \right> 
\nonumber\\[0.133ex]
=& \; - \frac{g^2 \, \vec B^{\,2}}{4 m_r}
= - \epsilon_0 \, (\hbar \, c)^3 \, \frac{g^2 \, \vec B^{\,2}}{4 m_r} \,,
\end{align}
where $m_r$ is the reduced mass of the bound system
and SI mksA units are restored in the last step.
Otherwise, according to Table 5 of Ref.~\cite{EhEtAl2009},
we have
\begin{equation}
g < 4.9 \times 10^{-7} \, {\rm GeV} \,,
\qquad
m_\phi \lesssim 0.5 \, {\rm meV} \,.
\end{equation}
For the parameters $| \vec B | = 5 \, {\rm T}$ and 
$g = 5 \times 10^{-7} {\rm GeV}^{-1}$, we obtain
\begin{equation}
\delta E_H = -1.67 \times 10^{-31} \, {\rm eV} \,,
\qquad
\delta E_{\mu H} = -6.28 \times 10^{-34} \, {\rm eV} \,.
\end{equation}
The smallness of these results excludes ALPs as possible
explanations for the proton radius puzzle.
A possible scenario with a nonvanishing vacuum expectation 
value of the axion field (see also Refs.~\cite{DuvB2009,BaCMDe2014})
is studied in Appendix~\ref{appa}.

%
%
\section{Strong--Field Electrodynamics}
\label{sec3}

Muonic bound systems have been used as probes of the 
strongest electromagnetic fields since the 1970s
(see Ref.~\cite{BrMo1978}), but progress was eventually 
hindered due to electron screening~\cite{BeEtAl1986}.
Typically, transitions in high-$Z$ muonic ions 
involve highly excited, non-$S$ 
states~\cite{DeKeSaHe1980,RuEtAl1984,OfEtAl1991},
where the average field experienced by the orbiting 
electron is reduced due to the higher principle 
quantum number. In view of the current muonic hydrogen discrepancy,
it is useful to recall just how strong these fields
are, especially in very simple bound systems,
where shielding electrons are absent~\cite{KaHi2012,UmJo2014}.
The conceivable presence of novel phenomena
in the very strong electromagnetic fields 
within highly charged ions has been mentioned as
a significant motivation for the study of these
systems~\cite{IoEtAl1988,MoPlSo1998,SoEtAl1998,GuEtAl2005}.
According to Eq.~(2) of Ref.~\cite{SoEtAl1998}
and the more comprehensive discussion of Ref.~\cite{IoEtAl1988},
a conceivable nonlinear correction term 
(contact interaction) has been mentioned 
for high-field quantum electrodynamics.
In view of this situation, it is indicated to 
compare the field strengths in highly charged 
(electronic) ions to those reached for low excited
states in muonic hydrogen and low-$Z$ muonic ions.

A measure for the strongest electromagnetic fields
that can be described by perturbative electrodynamics is 
the Schwinger critical field 
strength~\cite{Sa1931a,Sa1931b,HeEu1936,Sc1951}
\begin{equation}
\label{Ecr}
E_{\rm cr} = 1.32 \times 10^{18} \, \frac{{\rm V}}{{\rm m}} \,.
\end{equation}
The electric field around the proton reaches the Schwinger critical 
field already at a distance $0.116 \, a_\mu$ Bohr radii of the 
muonic hydrogen system, where $a_\mu$ is given in 
Eq.~\eqref{amu}.
Let us consider bound one-muon ions in the region
of low nuclear charge numbers $1 \leq Z \leq 5$.
The probability of finding a $1S$ muon inside the 
region of super-critical field strength,
in one-muon ions of nuclear charge number
$1 \leq Z \leq 5$, is evaluated as follows,
\begin{subequations}
\begin{align}
p_{\rm cr}(Z=1) =& \; 0.17 \, \% \,, \\[0.133ex]
p_{\rm cr}(Z=2) =& \; 1.18 \, \% \,, \\[0.133ex]
p_{\rm cr}(Z=3) =& \; 3.36 \, \% \,, \\[0.133ex]
p_{\rm cr}(Z=4) =& \; 6.73 \, \% \,, \\[0.133ex]
p_{\rm cr}(Z=5) =& \; 11.2 \, \% \,.
\end{align}
\end{subequations}
The field scales as $1/r^2$ for small distances.
In Fig.~\ref{fig3}, 
to supplement a corresponding investigation in
Fig.~2 of Ref.~\cite{SoEtAl1998},
we investigate the electric field strength 
experienced by a bound
muon in a muonic hydrogenlike system (only one orbiting particle) in the
region of low nuclear charge number. 
We start from the ground-state expectation value
of the electric-field operator, which is obtained as the 
gradient of the Coulomb potential. 
Within the 
nonrelativistic approximation
(in SI mksA units), the result reads 
\begin{subequations}
\label{std}
\begin{align}
\label{stda}
& \langle E \rangle  
= \left< 1S \left| \left( - \frac{\partial}{\partial r} 
\frac{Z |e|}{4 \pi \epsilon_0 \, r} \right) \right| 1S \right> =
2 \, Z^3 \, \frac{m_r^2}{m_e^2} \, {\mathcal E}_0 \,, \\[0.133ex]
\label{stdb}
& {\mathcal E}_0 =
\frac{e \, \alpha_{\rm QED}^2 \, m_e^2 \, c^2}%
{4 \pi \, \epsilon_0 \, \hbar^2} =
5.14 \times 10^{11} \, \frac{{\rm V}}{{\rm m}} \,.
\end{align}
\end{subequations}
Here, $m_r$ is the reduced mass of the atomic 
system, $m_e$ is the electron mass, 
and ${\mathcal E}_0 $ denotes the ``standard'' atomic field 
strength observed at one Bohr radius in a 
standard hydrogen atom (it is equal to the 
atomic unit of the electric field strength).
The prefactor $2$~in Eq.~\eqref{stda} is a consequence
of our taking the quantum mechanical expectation value
as opposed to evaluating the classical expression
at the (shifted) Bohr radius.
For ultra-relativistic systems, Eq.~\eqref{stda} 
is replaced by the expectation value of the 
fully relativistic Dirac--Coulomb wave function~\cite{SwDr1991a};
the relativistic correction factor amounts to the 
replacement
\begin{equation}
\langle E \rangle \mapsto 
\frac{\langle E \rangle}{2 - \sqrt{1 - (Z\alpha_{\rm QED})^2} - 2 \, (Z\alpha_{\rm QED})^2} \,,
\end{equation}
which does not change the order-of-magnitude of the result.
The decisive factor in Eq.~\eqref{stda} is the
prefactor $Z^3 \, (m_r/m_e)^2$, which is responsible for an 
enhancement of the field strength by six orders of 
magnitude in the range $1 \leq Z \leq 92$ for the electronic system,
but also for a considerable enhancement in muonic systems,
where 
\begin{equation}
\left( \frac{m_r}{m_e} \right)^2 \to
\left( \frac{m_\mu \, m_p}{(m_\mu + m_p) \, m_e} \right)^2 \approx
3.45 \times 10^4 \,.
\end{equation}
For a one-muon ion, the average electric 
field strengths at $Z = 4$ and $Z = 5$ surpass
the average electric field strength in hydrogenlike Uranium
(see Fig.~\ref{fig3}).

Furthermore, the average field strength
experienced by a bound $1S$ electron in one-muon ions
with $Z=4$ and $Z=5$ is given as 
\begin{subequations}
\begin{align}
\langle E \rangle_{\mu,Z=4} =& \; 1.72 \, E_{\rm cr} \,,
\\[0.133ex]
\langle E \rangle_{\mu,Z=5} =& \; 3.36 \, E_{\rm cr} \,,
\end{align}
\end{subequations}
thus surpassing (in terms of quantum mechanical average)
the Schwinger critical field strength.

The HERCULES laser~\cite{YaEtAl2008}
(still) sets the standard for the highest achievable
laser intensities to date, with a peak intensity of about 
$2 \times 10^{22} \, {\rm W} \, {\rm cm}^{-2}$.
In the future, such facilities are supposed to 
reach intensities in the range
$10^{23}$---$10^{24} \, {\rm W} \, {\rm cm}^{-2}$.
An intensity of $10^{24} \, {\rm W} \, {\rm cm}^{-2}$
corresponds to an electric field strength of 
\begin{equation}
E_L = 2.74 \times 10^{15} \, \frac{{\rm V}}{{\rm m}} \,,
\end{equation}
which is surpassed in the muonic system ($1 \leq Z \leq 5$)
as well as medium-$Z$ and high-$Z$ bound 
quantum electrodynamic (QED) systems (with $Z \geq 14$, see Fig.~\ref{fig3}).
It is thus evident that bound muonic systems
offer a competing alternative to the exploration 
of the strong-field QED regime,
complementary to strong laser systems~\cite{Ke2001}.

One might argue that the time average of the oscillating laser fields is
zero, as much as the spatial (vector) average of the electric field (vector),
taken over the spherically symmetric $S$ wave function, vanishes. However,
the exploration of the strong-field domain of electrodynamics is not
precluded by the oscillating or 
spherically symmetric nature of the fields.  
One easily estimates that the (fluctuating) electric fields
inside the proton, given the fact that the three valence quarks cannot be
further apart than $0.8\,{\rm fm}$, are of order $E_p \sim 10^{21} \,
\frac{{\rm V}}{{\rm m}}$ and thus exceed the Schwinger critical field
strength $E_{\rm cr}$ of about $E_{\rm cr} = 1.32 \times 10^{18} \,
\frac{{\rm V}}{{\rm m}}$ by three orders of magnitude. 
Conceivable corrections to the muonic hydrogen 
spectrum due to the high field strengths have been
discussed in Refs.~\cite{Je2013pra,PaMe2014,Mi2015}.
Just to avoid a misunderstanding, we should clarify that the recently discussed
hypothesis of nonperturbative lepton pairs inside the
proton~\cite{Je2013pra,PaMe2014,Mi2015} certainly does not imply the production
of such pairs from the vacuum inside the nucleus; the vacuum is known to
``spark'' only if the critical field strength is maintained over a sufficiently
large space-time interval which is absent in muonic hydrogen. The hypothesis
discussed in Refs.~\cite{Je2013pra,PaMe2014,Mi2015} merely implies that the
highly nonperturbative nature of strong interactions (quantum chromodynamics)
inside the proton, which involves electrically charged constituent as well as
sea quarks, might lead to effective lepton-proton interactions which have so
far been overlooked in theoretical treatments (see
Refs.~\cite{Je2013pra,PaMe2014,Mi2015} and Appendix~\ref{appb}).


Finally, a remark on the relationship of the light muonic 
systems and the strong electric fields to the 
``classical'' strong-field systems (highly charged ions)
is in order. In these latter systems, the (initially positive-energy) 
$1S$ level can be shown to approach the negative 
continuum, effectively ``sparking'' the vacuum~\cite{ZePo1971,RaMuGr1974}.
A single proton of course is unable to create 
such an effect, but the proximity of the bound muon
to the proton (nucleus) generates the extreme fields 
and the corresponding quantum mechanical expectation values
that contribute to the interest in muonic bound systems.

%
%
\section{Non--Resonant Effects and Transition Frequencies}
\label{sec4}

Discrepancies of Lamb shift experiments and theory have been explored for a
long time.  For example, a rather well-known accurate Lamb shift experiment in
helium~\cite{vWKwDr1991} had long been in disagreement with theory (the
discrepancy was resolved in Refs.~\cite{vWHoDr2000,JeDr2004}).  A
measurement of the $^4$He nuclear radius using muonic helium ions is currently
in progress~\cite{PoPriv2014}.  In many cases, nuclear radius determinations
using electronic and muonic bound systems complement each
other~\cite{Je2011aop2}. One may add that additional experiments on electronic
helium ions (as opposed to muonic helium ions) would be able to shed additional
light on the ``generalized'' proton radius puzzle, or ``nuclear size effect
puzzle'', because they would enable us to compare the ``electronically
measured'' radius of $^4$He with the ``muonically measured'' radius; a
corresponding experimental setup was recently proposed~\cite{HeEtAl2009}.
In particular, it would be rather interesting to compare the ``anisotropy
method'' used in Refs.~\cite{vWKwDr1991,vWHoDr2000} with other spectroscopic
techniques.

Historical developments encourage us to search for additional conceivable
explanations of the proton radius puzzle in systematic effects that may not
have been fully appreciated in even the most carefully planned experiments.
One such set of corrections is given by so-called off-resonant corrections to
frequency measurements. 
In Ref.~\cite{CaRaSt1982}, it was stressed that
an accurate understanding of the line shape of quantum transitions to
neighboring levels can lead to surprising phenomena such as prevention of
fluorescence; for precision experiments, this finding highlights the
necessity of including a good line-shape model.  Because the non-resonant
corrections to the line shape involve mixed products of dipole operators
connecting the resonant and off-resonant levels, these effects are also
referred to as ``cross-damping'' terms in quantum
optics~\cite{FiSw2002,FiSw2004}
[see also Eq.~(9) of Ref.~\cite{JeMo2002}].  In Sec.~III of Ref.~\cite{JeMo2002} 
[see the text after Eq.~(15) {\em ibid.}], 
the authors investigate off-resonant
effects in differential as opposed to angular-averaged cross sections. 
Quantum interference effects can be 
excluded as an explanation of the proton radius 
discrepancy in muonic systems~\cite{AmEtAl2015},
mainly because the proton radius discrepancy, converted 
to frequency units, is much larger than the 
natural linewidth of the transitions in the muonic systems.
However, the situation is different for atomic hydrogen, 
where spectral lines have to be split to much higher relative accuracy.
In order to gauge possible concomitant systematic shifts of the accurately measured
frequencies, especially those involving highly excited states of 
(atomic) hydrogen and
deuterium, improved measurements of hydrogen $2S$--$nP$ lines 
are currently being
pursued~\cite{UdPriv2014}, while an improved measurement of the
``classic'' $2S$--$2P_{1/2}$ Lamb shift is also 
planned~\cite{HePriv2014}.  Both of these experiments have the potential
of clarifying the ``electronic hydrogen'' side of the proton radius puzzle.

In order to understand the importance of the nonresonant terms, and see if
they can potentially contribute to the explanation of the proton radius
puzzle, let us recall that a typical nonresonant energy shift due to
neighboring levels, still displaced by an energy shift $\Delta E_n$
commensurate with a change in the principal quantum number,
is~\cite{JeMo2002,Lo1952}
\begin{equation}
\label{low}
\delta E = \frac{(\hbar \, \Gamma)^2}{\Delta E_n} \sim
\alpha_{\rm QED}^8 \, m_e \, c^2 \,,
\end{equation}
where $\Gamma$ is the decay width of the reference state
and the term after the ``$\sim$'' sign is a parametric 
estimate according to the $Z\alpha_{\rm QED}$-expansion~\cite{BeSa1957}.
The shift~\eqref{low}, which according to Low~\cite{Lo1952} defines
the ultimate limit to which energy levels 
can be resolved in spectroscopic experiments, is too small to explain the 
proton radius puzzle 
(we have $\hbar \, \Gamma \sim \alpha_{\rm QED}^5 \, m_e \, c^2$,
while $\Delta E_n \sim \alpha_{\rm QED}^2 \, m_e \, c^2$
for a transition with a change in the principal
quantum number). By contrast, in differential cross sections, the 
shift due to neighboring levels removed
only by the fine-structure is proportional to~\cite{JeMo2002}
\begin{equation}
\label{high}
\delta E = \frac{(\hbar \, \Gamma)^2}{\Delta E_{\rm fs}} \sim
\alpha_{\rm QED}^6 \, m_e \, c^2 \,,
\end{equation}
where $\Delta E_{\rm fs}\sim  \alpha_{\rm QED}^4 \, m_e \, c^2$ is 
of the order of a typical fine-structure interval.
According to Eqs.~(9) and~(12) of Ref.~\cite{JeMo2002},
there is an additional prefactor $1/2$ to consider for the 
shift of the center of the half-maximum values of the 
resonance curve, while this prefactor is $1/4$ for 
the Lorentzian maximum itself. 
The presence of this additional prefactor has no effect
on the phenomenological significance of the estimates 
to be discussed in the following.  The shift given in Eq.~\eqref{high} 
is of sufficient magnitude to explain the
muonic hydrogen discrepancy.

Let us perform some order-of-magnitude estimate 
to explore the possibility of explaining the proton 
radius puzzle on the basis of non-resonant corrections.
The reduced electron Compton wavelength is 
$\lambdabar_C = \hbar/(m_e \, c) =
386.159 \, {\rm fm}$.
The ratio of $\lambdabar_C$ to the proton radius,
which we assume to be given by 
$r_p \approx 0.88 \,  {\rm fm}$, is given as
\begin{equation}
\xi = \frac{r_p}{\lambdabar_C} = 2.27 \times 10^{-3} \,.
\end{equation}
According to Eq.~(51) of Ref.~\cite{MoTaNe2008}
(see also Table~10 of Ref.~\cite{EiGrSh2001}),
the leading-order finite-size effect for the $2S$ state is 
(non-recoil limit),
\begin{equation}
E_{\rm FS} 
= \frac{1}{12} \, (Z\alpha)^4 \, m_e c^2 \, \xi^2 
\approx 150 \, {\rm kHz} .
\end{equation}
We defined the ``proton puzzle prefactor'' 
$\chi_{\rm PP}$ as 
\begin{equation}
\chi_{\rm PP} = \frac{0.88^2 - 0.84^2}{0.88^2} = 0.089 \,,
\end{equation}
leading to a ``proton puzzle energy 
shift'' $E_{\rm PP}$ for the $2S$ state of
\begin{equation}
E_{\rm PP} = \chi_{\rm PP} \, E_{\rm FS} 
\approx 13 \, {\rm kHz} .
\end{equation}
We aim to investigate the possible presence of 
significant off-resonant corrections to the 
$2S$--$4P_{1/2}$ and 
$2S$--$4P_{3/2}$ frequencies~\cite{BeHiBo1995},
as well as $2S$--$8D_{3/2}$ and 
$2S$--$8D_{5/2}$ frequencies~\cite{BeEtAl1997},
and $2S$--$12D$ transitions~\cite{ScEtAl1999}.
To this end, we first recall that 
the fine-structure energy difference, for $P$ 
and $D$ states in hydrogen, is 
\begin{subequations}
\label{chiF}
\begin{align}
\calF_{nP} = & \; E_{nP_{3/2}} - E_{nP_{1/2}} =
\chi_{\calF P} \, \frac{(Z\alpha)^4 \, m_e \, c^2}{n^3} \,,
\\[0.133ex]
\calF_{nD} = & \; E_{nD_{5/2}} - E_{nD_{3/2}} =
\chi_{\calF D} \, \frac{(Z \alpha)^4 \, m_e \, c^2}{n^3} \,,
\\[0.133ex]
\chi_{\calF P} =& \; \frac14 \,,
\qquad
\qquad \chi_{\calF D}  = \frac{1}{12} \,.
\end{align}
\end{subequations}
According to p.~266 of Ref.~\cite{BeSa1957},
the one-photon decay width of $nP$ and $nD$ states can be estimated as
(independent of the total angular momentum)
\begin{subequations}
\label{chiGamma}
\begin{align}
\Gamma_{nP} \approx & \; 
\chi_{\Gamma P} \, \frac{\alpha \, (Z\alpha)^4 m_e c^2}{\hbar n^3} \,,
\qquad
\chi_{\Gamma P} = 0.311 \,,
\\[0.133ex]
\Gamma_{nD} \approx & \;
\chi_{\Gamma D} \, \frac{\alpha \, (Z\alpha)^4 m_e c^2}{\hbar n^3} \,,
\qquad
\chi_{\Gamma D} = 0.107 \,.
\end{align}
\end{subequations}

We now focus on a potential nonresonant correction to 
the transition frequencies, 
due to neighboring fine-structure levels.
This choice is motivated in part by a remark 
in the text in the right-hand column of
the second page of Ref.~\cite{BeEtAl1997},
where it is confirmed that neighboring
hyperfine structure levels are taken into account in the line-shape 
model used in Ref.~\cite{BeEtAl1997} 
(but those levels displaced by the fine structure 
apparently are not taken into account).

According to Ref.~\cite{JeMo2002},
in an angle-differential cross section, the off-resonant shift 
due to neighboring fine-structure levels 
can be estimated as follows. For a $2S$--$nP$ transition,
\begin{equation}
E_{\rm OR} = \frac{(\hbar \Gamma)^2_{n}}{\calF_{n}} =
\frac{\chi_{\Gamma}^2}{\chi_{\calF}} \,
\frac{\alpha^2 \, (Z\alpha)^4 m_e c^2}{n^3} \,,
\end{equation}
where one has to replace the prefactors as 
$\chi_{\Gamma} \to \chi_{\Gamma P,D}$
and $\chi_{\calF} \to \chi_{\calF P,D}$, respectively,
according to the estimates given 
in Eqs.~\eqref{chiF} and~\eqref{chiGamma}.

It is interesting to investigate 
the ratio of the proton size puzzle energy shift to the 
natural linewidth as a measure of how precisely the line 
has to be split in order to resolve the proton size puzzle.
It is given as ($2S$-$nP$ transitions),
\begin{equation}
R_P = \frac{E_{\rm PP}}{\hbar \Gamma_{nP}}
= \frac{n^3 \, \chi_{\rm PP} \, \xi^2}{12 \, \alpha \, \chi_{\Gamma P}}
= 1.68 \times 10^{-5} \, n^3 \,.
\end{equation}
Example values for $2S$-$nP$ are $R_P(n=4) = 0.0011$,
$R_P(n=8) = 0.008$, and $R_P(n=12) = 0.029$.
So, in order to resolve the proton size puzzle 
based on the $2S$-$4P$ transition, one has to understand the line width 
to better than 1 part in 1000.
The work in~\cite{BeHiBo1995} reaches an accuracy close to this 
limit: The experimental accuracy for the $2S$-$4P$ transitions
is on the order of $\sim 12 \, {\rm kHz}$, 
to be compared to a natural line width 
of $\sim 13 \, {\rm MHz}$.
The ratio $R_P$ becomes significantly more favorable
for transitions to higher excited $P$ states.

The corresponding estimate for $2S$-$nD$ transitions is
\begin{equation}
R_D = \frac{E_{\rm PP}}{\hbar \Gamma_{nD}}
= \frac{n^3 \, \chi_{\rm PP} \, \xi^2}{12 \, \alpha \, \chi_{\Gamma D}}
= 4.86 \times 10^{-5} \, n^3 \,.
\end{equation}
For $2S$--$nD$ transitions with $n=4,8,12$, we have
$R_D(n=4) = 0.0031$, $R_D(n=8) = 0.025$,
and $R_D(n=12) = 0.084$.
It means that in order to resolve the proton size puzzle
based on the $2S$-$12D$ transition~\cite{ScEtAl1999}, 
one has to understand the line width
only to (roughly) 1 part in 12.

Another interesting quantity is the ratio of the 
off-resonant terms to the natural linewidth.
It measures how accurately the natural line width has 
to be split in order to see the off-resonant effects.
For $2S$--$nP$ transitions and 
$2S$--$nD$ transitions, it is given by 
\begin{subequations}
\begin{align}
S_P =& \; \frac{E_{\rm OR}}{\hbar \Gamma_{nP}}
= \frac{\alpha \, \chi_{\Gamma P}}{\chi_{\calF D}}
\approx \frac{1}{110} \,,
\\[0.133ex]
S_D =& \; \frac{E_{\rm OR}}{\hbar \Gamma_{nD}}
= \frac{\alpha \, \chi_{\Gamma D}}{\chi_{\calF D}}
\approx \frac{1}{106} \,,
\end{align}
\end{subequations}
independent of $n$.
It is also very important to compare the 
``proton size puzzle energy shift'' to the 
off-resonant shift. It is given by 
($2S$--$nP$ transitions)
\begin{equation}
T_P = \frac{E_{\rm PP}}{E_{\rm OR}} =
\frac{R_P}{S_P} 
= \frac{n^3 \, \chi_{\rm PP} \, \chi_{\calF P} \, \xi^2 }%
{12 \, \alpha^2 \, \chi_{\Gamma P} }
=  1.85 \times 10^{-3} \, n^3 \,.
\end{equation}
For the $2S$--$4P$ transition, one has $T_P(n=4) = 0.118$,
implying that the off-resonant, cross-damping shift
due to neighboring fine-structure levels 
is roughly ten times larger than the 
energy shift corresponding to the 
proton size puzzle for the $2S$~level.
We conclude that, unless one uses an appropriate 
$4 \pi$ detector to eliminate the nonresonant terms,
one has to understand the line shape of the 
$2S$--$4P$ transition extremely well
in order to resolve proton radius puzzle
based on this transition. From a complementary viewpoint, the 
line shape of the $2S$--$4P$ transition could be an 
an excellent tool for studying the nonresonant 
cross-damping terms.

For $2S$-$nD$ transitions, we have
\begin{equation}
T_D = \frac{E_{\rm PP}}{E_{\rm OR}} =
\frac{R_D}{S_D} 
= \frac{n^3 \, \chi_{\rm PP} \, \chi_{\calF D} \, \xi^2 }%
{12 \, \alpha^2 \, \chi_{\Gamma D} }
=  5.16 \times 10^{-3} \, n^3 \,.
\end{equation}
Examples are $T_D(n=8) = 2.64$, and $T_D(n=12) = 8.92$.  For the $2S$--$8D$
transitions and $2S$--$12D$ transitions studied in Refs.~\cite{BeEtAl1997}
and~\cite{ScEtAl1999}, respectively, this means that the estimated ratio of
the proton size puzzle energy shift to the off-resonant contribution is
larger than unity.  One could thus tentatively conclude that the inclusion of
any conceivable nonresonant corrections is not likely to shift the
experimental results reported in Refs.~\cite{BeEtAl1997}
and~\cite{ScEtAl1999} on a level commensurate with the proton radius puzzle
energy shift.

In summary, our estimates would suggest that $2S$--$nD$ transitions to highly
excited $D$ states provide for the most favorable ``signal-to-noise'' ratio
$E_{\rm PP}/E_{\rm OR}$ [ratio of proton size puzzle energy shift to the off-resonant
energy shift, with $T_D(n=12)=8.92$]. 
In view of $R_D(n=12) = 0.084$, the proton puzzle energy shift
enters at about $1/12$ of the natural line width~\cite{ScEtAl1999} for $n=12$.
Because $S_D \approx 1/106$, the off-resonant terms are suppressed by about two
orders of magnitude in relation to the natural linewidth, which is smaller than
the proton radius puzzle energy shift by roughly another order of magnitude.
An inspection of Fig.~1 of Ref.~\cite{BeEtAl2013} (see also Fig.~1 of
Ref.~\cite{BeEtAl2013jpconf}) would indicate that the $2S$--$8D$ and
$2S$--$12D$ transitions are consistent with a proton radius, derived from
hydrogen experiments, which is significantly larger than the muonic hydrogen
result.  A least-squares analysis of all accurately measured hydrogen
transitions yields the proton radius $r_p = 0.8802(80) \, {\rm fm}$ (see
Table~XLV of Ref.~\cite{MoTaNe2012}).  For comparison, we have exclusively
taken the data from the $2S$--$8D$ and $2S$--$12D$ transitions reported in
Refs.~\cite{BeEtAl1997,ScEtAl1999}, together with the latest $1S$--$2S$
result~\cite{MaEtAl2013prl}, and current theory as summarized in
Refs.~\cite{MoTaNe2008,MoTaNe2012}, and calculated a naive statistical average
of the proton radii derived from $2S$--$8D$ and $2S$--$12D$ transitions
(disregarding the covariances among the data which otherwise leads to a much
more accurate value for the proton radius~\cite{JeKoLBMoTa2005}).  With this
approach, the result
from $2S$--$8D$ and $2S$--$12D$ transitions alone is $r_p = 0.873(17) \, {\rm
fm}$, still larger than the muonic hydrogen value by $2\sigma$.  While the
reconsideration of cross-damping terms for hydrogen transitions would be very
helpful in clarifying a conceivable contribution to the solution of the proton
size puzzle, our estimates suggest that it would be very surprising if the
proton size puzzle were to find a full explanation based on the cross-damping
terms alone.  The off-resonant terms seem to be most effectively suppressed in
transitions to highly excited $D$ states.

%
%
\section{Conclusions}
\label{sec5}

In this paper, we explored the remaining options for the explanation of 
the persistent proton radius discrepancy~\cite{PoEtAl2010,AnEtAl2013}.
Specifically, in Sec.~\ref{sec2},
we supplemented previous attempts to find an explanation for the proton radius
puzzle based on ``subversive'' virtual particles; all of these appear to
require fine-tuning of the coupling constants and no compelling set of
quantum numbers has as yet been found for the virtual particle which could
potentially explain the discrepancy of theory and experiment in (muonic)
hydrogen within the limits set by other precision experiments such as the
electron and muon $g$ factors.  Virtual particle explanations appear to be
disfavored at the current stage, and other models depend on rather drastic
hypotheses such as symmetry breaking terms that affect only the muon sector
of the Standard Model (but not electrons or positrons). 
Here, we supplemented the discussions of virtual particles
by a calculation of the effective potential describing the 
leading correction to the Coulomb interaction due to axion--photon
conversion in the (strong) magnetic fields used in the 
muonic hydrogen experiments~\cite{PoEtAl2010,AnEtAl2013}.

In Sec.~\ref{sec3}, we continued to explore the
typical electric fields in a low-$Z$ bound muonic system.  These fields are
seen to be commensurate with, or even exceed the Schwinger critical field
strength.  Because of the lack of electron screening, the one-muon ions can
be interpreted as the most sensitive probes of high-field physics to date.
The hypothesis of nonperturbative lepton 
pairs inside the proton and their conceivable influence
on electron-proton and muon-proton interactions 
(see Refs.~\cite{Je2013pra,PaMe2014,Mi2015}) is based on the 
interplay of nonperturbative quantum chromodynamics
with quantum electrodynamics (see Appendix~\ref{appb}). 
A breakdown of perturbative quantum electrodynamics
is not necessary for the existence of the conjectured effect~\cite{Mi2015}.
Muon-proton scattering experiments will be an important
cornerstone in the further clarification 
of the electron-muon universality in lepton-proton 
interactions (MUSE collaboration, see
Ref.~\cite{KoEtAl2014}). 

Finally, in Sec.~\ref{sec4}, the role of
nonresonant line shifts in differential as opposed to total cross sections was
mentioned.  Two ongoing experimental efforts~\cite{UdPriv2014,HePriv2014}
share the aim of analyzing the process-dependent line shifts~\cite{Lo1952}
further. Transitions to highly excited $D$ states ($2S$--$nD$ transitions)
in hydrogen are identified in terms of favorable parameters for the 
suppression of nonresonant correction terms (cross-damping terms),
which otherwise could account for hitherto unexplored systematic
effects in atomic hydrogen experiments.
For the muonic hydrogen experiments, in contrast to the
experiments on ordinary hydrogen, it is not necessary to ``split'' the
resonance line in order to make the proton radius puzzle manifest; the
discrepancy is much larger than the width of the resonance line itself (see
Fig.~5 of Ref.~\cite{PoEtAl2010}).

The binding field strengths in muonic ions exceed those achievable in 
current and projected high-power laser systems.
The benefit of the low-$Z$ muonic ions produced in the 
high-intensity muon beams at the Paul--Scherrer--Institute
(PSI) lies in the ``clean'' environment provided by the 
one-muon ions, where all other bound electrons have been 
stripped and the interaction of the muon and the nucleus can be 
investigated spectroscopically to high accuracy.
From a theoretical point of view,
it appears to be hard to shed any further light 
on the proton radius puzzle without significant further 
stimulation from additional experimental 
spectroscopic or scattering data.

%
%
\begin{acknowledgments}
Helpful conversation with I.~N\'{a}ndori and M.~M.~Bush are gratefully
acknowledged.  The author is grateful to the Mainz Institute for Theoretical
Physics (MITP) for its hospitality and its partial support during the
completion of this work.  This research was supported by the National
Science Foundation (Grants PHY--1068547 and PHY--1403973).
\end{acknowledgments}

\appendix

%
%
\section{Heisenberg--Euler Lagrangian and Variational Calculus}
\label{appa}

In many cases, the leading perturbation to the 
Coulomb potential due to a new interaction
can be obtained by variational calculus;
we illustrate the procedure here,
on the basis of the Wichmann--Kroll correction to the Coulomb potential.
The Maxwell Lagrangian with the Heisenberg--Euler Lagrangian
reads as
(we switch to natural units with $\hbar = c = \epsilon_0 = 1$)
\begin{align}
\label{LeffSOURCE}
\calL =& \; 
\tfrac{1}{2} \, \left( \vec{E}^2 - \vec{B}^2 \right) 
\nonumber\\[0.133ex]
& \; + \frac{2 \, \alpha_{\rm QED}^2}{45 m^4} 
\left( \vec{E}^2 - \vec{B}^2 \right)^2  +
\frac{14 \, \alpha_{\rm QED}^2}{45 m^4} 
\left( \vec{E} \cdot \vec{B} \right)^2 \,.
\end{align}
If $\vec E$ is given by the gradient of a Coulomb field
and the magnetic field vanishes ($\vec B = \vec 0$), 
then $\calL$ is redefined to the expression
\begin{equation}
\calL = \frac{1}{2} \, \left( \vec{\nabla} \Phi \right)^2 +
\frac{2 \, \alpha_{\rm QED}^2}{45 m^4} \,
\left( \vec{\nabla} \Phi \right)^4 - \rho \, \Phi \,,
\end{equation}
where we add the source term.  In view of the relations
\begin{equation}
\frac{\partial \calL}{\partial \vec\nabla \Phi} = 
\vec{\nabla} \Phi 
+ \frac{8 \, \alpha_{\rm QED}^2}{45 m^4} \, \, \vec\nabla \Phi \, 
\left( \vec\nabla \Phi \right)^2 \,,
\qquad
\frac{ \partial \calL}{\partial \Phi} = -\rho \,,
\end{equation}
the variational equation 
$\vec\nabla \cdot \frac{\partial \calL}{\partial \vec\nabla \Phi} = 
\frac{ \partial \calL}{\partial \Phi}$ becomes
\begin{equation}
\vec{\nabla}^2 \Phi 
+ \frac{8 \, \alpha_{\rm QED}^2}{45 m^4}\, 
\left( \vec\nabla^2 \Phi \, \left( \vec\nabla \Phi \right)^2 
+ \vec\nabla \Phi \cdot \vec\nabla \left( \vec\nabla \Phi \right)^2 
\right) = -\rho  \,,
\end{equation}
which can be reformulated as 
\begin{equation}
\vec{\nabla}^2 \Phi + \frac{8 \, \alpha_{\rm QED}^2}{45 m^4} \, 
\left( \partial_r + \frac{2}{r} \right) \, \left( \partial_r \Phi \right)^3
= -\rho \,.
\end{equation}
where we assume that $\Phi$ is radially symmetric.
We set $\Phi = \Phi_C + \Xi$ where
$\Phi_C$ is the Coulomb potential and $\Xi$ is a quantum correction.
The charge density of the nucleus and the Coulomb potential are given by 
\begin{equation}
\rho(\vec r) = Z \, |e| \, \delta^{(3)}(\vec r) \,,
\qquad
\Phi_C(\vec r) = \frac{Z \, |e|}{4 \pi \, r} \,,
\end{equation}
where $\vec\nabla^2 \, \Phi_C(\vec r) = -\rho(\vec r)$,
so that, to first order in $\Xi$,
\begin{equation}
\label{plus_xi}
\left( \partial_r^2 + \frac{2}{r} \, \partial_r \right) \Xi + 
\frac{8 \, \alpha_{\rm QED}^2}{45 m^4} \, 
\left( \partial_r + \frac{2}{r} \right) 
\left( \partial_r \Phi_C \right)^3 = 0 \,.
\end{equation}
It is straightforward to 
observe that Eq.~\eqref{plus_xi} is solved by a potential proportional to $r^{-5}$,
\begin{equation}
\label{asymp}
\Phi = \Phi_C + \Xi = \frac{Z \, |e|}{4 \pi \, r} \, 
\left( 1 - \frac{2}{225} \, \frac{\alpha_{\rm QED}}{\pi} \, 
\frac{(Z\alpha_{\rm QED})^2}{(m \, r)^4} \right) \,,
\end{equation}
This is equal to the long-distance tail of the 
Wichmann-Kroll potential~\cite{WiKr1954,WiKr1956},
which is relevant to a distance range $r \sim a_0$,
where $a_0$ is the Bohr radius;
we here confirm the result given in Appendix~III of 
Ref.~\cite{WiKr1956}.

A few remarks are in order.  Matrix elements of a term of order $(\alpha_{\rm
QED}/\pi) \, (Z\alpha_{\rm QED})^3/(m^4 \, r^5)$ [see Eq.~\eqref{asymp}]
generate an energy shift proportional to $\alpha_{\rm QED} \, (Z\alpha_{\rm
QED})^8 \, m$ in hydrogenlike systems. 
By contrast, the leading term in the Wichmann--Kroll
potential is otherwise proportional to a Dirac-$\delta$ function and generates
an energy shift of the order of $\alpha_{\rm QED} \, (Z\alpha_{\rm QED})^6 \, m$.
The latter term is given by the high-energy (short-distance) regime not covered
by our variational ansatz.  Namely, the atomic nucleus and the Coulomb potential
and its derivative, the Coulomb field, vary considerably on the length scale of
an electron Compton wavelength, which exceeds the ``operational parameters'' of
the Heisenberg--Euler Lagrangian, so the result~\eqref{asymp} cannot be used
for distances closer than an electron Compton wavelength, i.e., it fails in the
immediate vicinity of the nucleus.

One might wonder why the functional form of the long-distance tail ($1/r^5$ for
the Wichmann--Kroll potential) is different from the corresponding term for the
Uehling potential, which decays exponentially at large distances
(see~\cite{Je2011aop1} and references therein).  The answer to this question is
that the Wichmann--Kroll potential, which is generated by Feynman diagrams with
at least four electromagnetic interaction terms inside the loop, can be related
to the Heisenberg--Euler effective Lagrangian, which is valid for the
long-distance tail of the potential, while the corresponding term, for the
Uehling potential (with only two electromagnetic interaction terms inside the
loop) would otherwise generate a term proportional to $\vec E^2$ that is
absorbed in the $Z_3$ renormalization of the electromagnetic
charge~\cite{ItZu1980}. Hence, the tail of the Uehling potential decays
exponentially, akin to a Yukawa potential, with a range of the potential being
proportional to the electron Compton wavelength
(Sec.~2.4 of Ref.~\cite{Je2011aop1}).

After these intermediate considerations, we may proceed to 
apply our variational ansatz to a calculation of interest 
in the context of the subject matter of the current investigation.
Namely, for a nonvanishing vacuum expectation value 
$\left< \phi \right> \neq 0$ of the axion-like particle
as a dark matter candidate, the Lagrangian~\cite{DuvB2009,BaCMDe2014}
\begin{equation}
\label{LA}
\calL_A = \frac12 \, (\vec E^2 - \vec B^2)
+ g \, \left< \phi \right> \, \vec E \cdot \vec B
\end{equation}
is exact up to possible QED or axion loop corrections;
in contrast to the Heisenberg-Euler Lagrangian, it is 
not the result of ``integrating
out'' the fermionic degrees of freedom which limits 
the ``operational parameters'' of the 
Lagrangian~\eqref{LeffSOURCE}.
Hence, we are not at risk of exceeding the 
``operational parameters of the variational ansatz''
when we use the axion background Lagrangian~\eqref{LA} 
to calculate a possible correction
to the Coulomb potential due to dark matter physics.
If $\vec E = -\vec\nabla \Phi$ is generated 
by a (possibly distorted) Coulomb field and $\vec B$ 
is the (possibly inhomogeneous) external magnetic field,
then the Lagrange density $\calL_A$
is redefined to read (adding the source term $\rho \, \Phi$)
\begin{equation}
\calL_A = \frac{1}{2} \, \left( \vec{\nabla} \Phi \right)^2
- g \, \left< \phi \right> \, \vec B \cdot \vec\nabla \Phi
- \rho \, \Phi \,.
\end{equation}
The variational equation 
\begin{equation}
\label{variat}
\vec\nabla \cdot \frac{\partial \calL_A}{\partial \vec\nabla \Phi} =
\frac{ \partial \calL_A}{\partial \Phi}
\end{equation}
becomes
\begin{equation}
\vec{\nabla}^2 \Phi
- g \, \left< \phi \right> \, \vec\nabla \cdot \vec B
= -\rho \,.
\end{equation}
In the absence of magnetic monopoles, the leading 
correction to the Coulomb potential mediated by the axion 
vacuum expectation value 
thus vanishes, even for very strong and inhomogeneous external 
magnetic fields $\vec B$.

One more remark is in order.
The direct coupling of the fermion to the axion~\cite{PDG2012}
is of the derivative form $\calL_{Aff} = (C_f/(2 f_A)) \,
\overline\psi_f \, \gamma^\mu \, \gamma_5 \, \psi_f \,
\partial_\mu \phi$, where $C_f$ is a model-dependent 
constant. The Yukawa coupling is $g_{Aff} = C_f m_f/f_A$ 
and the ``fine-structure constant'' is $g^2_{Aff}/(4 \pi)$;
energy loss arguments from the SN1987A supernova typically
give bounds in the range $g^2_{Aff}/(4 \pi) \sim 10^{-21}$
(see Refs.~\cite{GrMaPe1989,Ra1990prep}).
This implies that a single axion exchange, or an
axion interaction insertion (e.g., into the fermion line of a
vacuum polarization diagram) suffers from a
suppression factor on the order of $g^2_{Aff}/(4 \pi) \sim 10^{-21}$
and is thus suppressed with respect to the corresponding 
photon exchange diagram (coupling parameter $\alpha_{\rm QED}$)
by roughly 18~orders of magnitude.
The fermion-axion coupling thus is too small to explain the 
proton radius puzzle.
Axion-mediated effects as well as weak interactions~\cite{Ei2012}
can thus also be excluded as possible explanations for the 
proton radius puzzle.

%
%
\section{Strong Fields in the Proton,\\
Interplay of QED and QCD}
\label{appb}

The presence of a very small fraction
of light sea fermions, conceivably due to 
a nonperturbative mechanism, inside the proton,
was recently mentioned in Refs.~\cite{Je2013pra,PaMe2014}.
One might counter-argue that the QED running coupling 
constant, at distances commensurate with the 
proton radius, still is small against unity,
and that this precludes
a nonperturbative mechanism leading to sea fermions 
inside the proton. In Sec.~III of Ref.~\cite{Je2013pra},
it is argued that the highly nonperturbative 
quantum chromodynamic (QCD) 
nature of a hypothetical electrically neutral proton
receives a correction due to 
the electroweak interactions, as they are switched back on,
and that, due to the highly nonlinear nonperturbative nature of QCD,
this reshaping can be much larger than the 
electromagnetic perturbation itself.

Alternatively, one might argue that the 
fundamentally nonperturbative nature of the 
QCD interaction inside the proton might leave room 
for effects that cannot be described by dispersion
relations and perturbation theory alone.
Namely, due to the nonperturbative nature of 
QCD, the three valence quarks of the proton are
supplemented, at any given time, by a large 
number of virtual sea quarks that 
emerge from the vacuum due to quantum corrections
to the gluon exchange~\cite{GaLa2009}. The sea quarks, as much 
as the valence quarks, are electrically charged,
off of their mass shell,
and may exchange photons. 
The propagator of these
photons, in turn, receives a correction due to 
vacuum polarization; hence, at any given time,
the proton wave function has a nonvanishing 
electron-positron content due to the light fermionic
vacuum bubbles. This is a persistent phenomenon 
because the quarks inside the proton are always 
highly virtual (off mass shell) in view of their
strong (mutual) interactions~\cite{Mi2015}. 

In Ref.~\cite{Mi2015}, the lepton pair content 
has recently been estimated based on a (perturbative) calculation
of the electron-positron vacuum polarization insertion 
into the radiative correction to a constituent 
quark's vector and axial vector current matrix elements.
According to Ref.~\cite{Je2013pra}, the 
virtual annihilation channel in positronium,
\begin{equation}
\label{B1}
\delta H = \frac{\pi \alpha_{\rm QED}}{2 m_e^2} \,
\left( 3 + \vec \sigma_+ \cdot \vec \sigma_-
\right)  \, \delta^3(r) \,,
\end{equation}
corresponds to an effective Hamiltonian for electron-proton interactions
of the form
\begin{equation}
\label{B2}
H_{\rm ann} = \epsilon_p \, \frac{3 \pi \alpha_{\rm QED}}{2 m_e^2} \, \delta^3(r) \,,
\end{equation}
where $\epsilon_p$ measures the electron-positron pair content inside the proton
and a value of $\epsilon_p = 2.1 \times 10^{-7}$ 
is found to be sufficient to explain the proton radius puzzle.
Near Eq.~(22) of Ref.~\cite{Mi2015}, it is argued that instead of
Eq.~\eqref{B1}, one should rather consider the Hamiltonian
\begin{equation}
\label{B3}
\delta H = \frac{\pi \alpha_{\rm QED}}{2 m_q^2} \,
\left( 3 + \vec \sigma_+ \cdot \vec \sigma_-
\right)  \, \delta^3(r) \,,
\end{equation}
where $m_q$ is a quark mass.  According to Ref.~\cite{Mi2015},
an appropriate choice is $m_q \approx 600 \, m_e$ 
(constituent value of one third of the mass of a proton).
Comparing Eqs.~\eqref{B1}---\eqref{B3}, one 
is led to the identification 
$\epsilon_p \sim m_e^2/m_q^2 \approx 2.8 \times 10^{-6}$ 
which is ``too large'' to explain the 
proton radius puzzle.
An estimate of the lepton pair content of the proton,
based on electron-positron vacuum polarization insertions
into the radiative correction to a constituent
quark's vector and axial vector current,
likewise leads to estimates for $\epsilon_p$ that are
much larger than the discrepancy.
According to Eqs.~(13) and~(21) of Ref.~\cite{Mi2015},
an estimate based on matrix elements of the 
current leads to values in the range 
\begin{equation}
\epsilon_p \sim 10 \, \left(
\frac{\alpha_{\rm QED}}{\pi} \right)^2 \sim
10^{-5} \gg 10^{-7} \,.
\end{equation}
Conversely, if one starts from Eq.~\eqref{B3} instead of~\eqref{B1},
arguing that the effective mass in the 
virtual annihilation
diagram should be the quark mass, and {\em additionally} invokes the 
suppression factor $\epsilon_p$ 
[see the text following Eq.~(22) of Ref.~\cite{Mi2015}],
then the resulting effect in muonic hydrogen becomes negligible
on the level of the proton radius discrepancy.
Guidance for the exploration of the conjectured 
sea lepton effect in future experiments
is given by the discussion surrounding Eq.~(23)
of Ref.~\cite{Mi2015}, where the functional dependence 
on the charge and mass numbers of the nucleus is discussed.

Nuclear structure corrections (nuclear polarizability corrections) are usually
taken into account with the use of dispersion relations~\cite{CaGoVa2014}.
This is certainly a valid approach for genuine excitations of the valence
quarks into excited states.  However, the light sea fermions are generated by a
QED correction to a nonperturbative process, namely, a correction to the
nonperturbative QCD interaction inside the proton; the latter gives rise to the
ubiquitous sea quarks.  Dispersion relations (Cutkosky rules) are available for
the treatment of the genuine excitations of the proton into its own excited
states, but it is unclear if the use of dispersion relations could capture the
effect of the sea fermions.  Because the sea quark interaction is
nonperturbative and the light fermion vacuum bubbles are inserted into the
photon exchange among the (nonperturbative) sea quarks, one does
not know where to cut the nonperturbative diagram and the dispersion relation
is not available.  For further details, we refer to
Refs.~\cite{Je2013pra,PaMe2014,Mi2015}.

\end{document}